\documentclass[twocolumn,twoside,slac]{revtex4}
\usepackage{graphicx}
\usepackage{fancyhdr}
\usepackage{textcomp}
\begin{document}

\title{BdbServer++: A User Driven Data Location and Retrieval Tool}

%

\author{A. Earl}
\affiliation{University of Edinburgh, Edinburgh EH9 3JZ, United Kingdom}
\author{A. Hasan}
\affiliation{Stanford Linear Accelerator Center, Stanford, California 94309}
\author{D. Boutigany}
\affiliation{Laboratoire de Physique des Particules, F-74941, Annecy-le-Vieux, France}

\begin{abstract}

\begin{center}
\normalsize\bf On behalf of the BaBar Computing Group
\end{center}

The adoption of Grid technology has the potential to greatly aid the BaBar experiment. BdbServer was originally designed to extract
copies of data from the Objectivity/DB database at SLAC and IN2P3. With data now stored in multiple locations in a variety of data
formats, we are enhancing this tool. This will enable users to extract selected deep copies of event collections and ship them to the
requested site using the facilities offered by the existing Grid infrastructure. By building on the work done by various groups in
BaBar, and the European DataGrid, we have successfully expanded the capabilities of the BdbServer software. This should provide a
framework for future work in data distribution.
\end{abstract}

\maketitle

\thispagestyle{fancy}


\section{Introduction}{\label{Introduction}}

BaBar, and HEP users in general, have long faced the challenge of locating and retrieving data remotely from large storage systems.  
With the increasingly distributed nature of the Computing Models used by HEP experiments there is an additional dispersed geographical
element to this problem. This paper outlines our understanding of the issues, with an emphasis on the BaBar experiment, and our 
attempts to address it. We outline the framework we have used and the projects progress.

The BdbServer++ project attempts to build on both prior work, the original BdbServer, and the current work of other groups, notably the
European DataGrid\cite{EDG} (EDG) and BaBarGrid\cite{BBG}. It is our intention to tackle the generic issues facing us in data
distribution, from primary to secondary and tertiary locations, and provide a useful tool to assist BaBar users. BdbServer++ is a
complimentary project to the work currently being done with the SDSC Storage Resource Broker\cite{WILCO} which, it is foreseen, will be
used by BaBar for bulk data transfer between primary computing sites.

In this paper we discuss the background to this problem, with specific reference to the evolving state of the BaBar Computing
Model\cite{CMWG}, the evolution and adoption of Grid technologies, and outline the areas which we believe have the greatest impact. 

We describe our work in this area and the results we have achieved so far. We discuss what has been learnt from this with the intention
of assisting other experiments in deciding how they wish to tackle the problem of data distribution for analysis and conclude with
suggestions for how our work will progress in the future. We comment on aspects of interest to the community.

In this paper Section \ref{BACKGROUND} provides a background to the project, covering the framework it has developed within,
the context and objectives. Section \ref{Progress} describes our current progress and sub-projects. Section \ref{Results} presents the
implementation issues of the project which are commented upon in context with other groups work in Section \ref{Related Work} and our
concluding remarks in Section \ref{Conclusions}.
   
\section{Background}\label{BACKGROUND}

\subsection{The BaBar Computing Model}

The BaBar Computing Model calls for there to be, at minimum, two distinct classes of Computing Site. These are currently entitled Tier
A and Tier C. Tier A sites are major resource centres, having MoUs with the Collaboration. Tier C sites refer to institute level
facilities, ranging from a single workstation to a major computing cluster and disk facility.

There are five locations currently classed as Tier A sites located in France\footnote{IN2P3, Lyon}, Germany\footnote{GridKa,
Karlsruhe}, Italy\footnote{INFN, Padova}, the UK\footnote{RAL, Didcot} and the USA\footnote{SLAC, Menlo Park}. These sites store a
quantity of Monte Carlo and Production data, provide computing resources and accounts for users. Sites specialise in specific tasks -
such as data storage, data re-processing and Monte Carlo production.

Due to the quantity of data produced by both the detector and simulation, no site stores a complete set of all data online. SLAC
maintains a complete archive but the majority of this is on Tape. The other sites may store all data in a particular format or a
selection of data. BaBar maintains data in a variety of formats, of different granularities. These range from the Mini\cite{MINI} to
the Micro\cite{MICRO} and Kanga\cite{KANGA}.

To summarise these formats, the Micro contained events with sufficient detail for most physics analyses and had a fast enough access
time to run a realistic analysis. However, it provided only the pion hypothesis track fit at the origin of the event which led to the
development of the Mini.

The Mini contains enough detail to provide a track reconstruction object which has information about fits for all physically distinct 
hypotheses valid throughout the entire detector. 

Kanga, the Kind ANd Gentle Analysis, is the de-facto format for physics analysis at Tier-C sites. It is used for micro analysis using
standard framework, but without Objectivity.

Tier C sites are the User Institutes, which exist in far greater numbers than the much larger Tier A sites. Tier C sites typically
request data be extracted from the database at a Tier A site and copied to their location for analysis. Data at these locations is
typically not available to users at other institutes.

\subsection{A Historical Context for BdbServer}

The BdbServer project was started in the late 1990's to automate the process of extracting collections from the Objectivity/DB
databases at SLAC and IN2P3. These sites have also behaved as the primary and backup database for the experiment. The presence of a
backup copy of the data at a different location serves as a safeguard against catastrophic problems at any one site.

The original BdbServer was developed using server resident Perl scripts and a generic system account. Users submitted requests to a
SLAC email address in a set format, specifying various parameters for the data they required. 

These requests were collated every 15 minutes by the BdbServer software and submitted to the batch system. The required data was
extracted and stored in temporary disk space at SLAC.  Users were informed by email when their request had been fulfilled.

As BdbServer was run as a single instance, and ran in a sequential manner over the requests it received, there were obvious issues of 
scalability and performance. The system lacked the ability  to remember previous requests and as a result was unable to utilise caching 
techniques for optimisation. 

The BdbServer++ project was started in 2002 to investigate migrating BdbServer to a Grid framework and to investigate options for
alternative interfaces. This was inspired by both the roll out of Grid infrastructure at Tier A sites and changes to the BaBar
Computing Model.

\subsection{BdbServer++ Objectives}

The primary aim of BdbServer++ is to migrate the currently available functionality of BdbServer to the existing Grid infrastructure.  
This should allow users to submit a request from a remote location to a central resource. This will extract a copy of the required data
and store it in a location accessible to the user.
 
The secondary aims of BdbServer++ are to improve the availability of Grid clients at user institutes, review available alternative
interfaces to this resource and investigate Grid technologies which could be of use to the project. 

As a byproduct we hope to increase the reliability, scalability and optimisation of the software.

\section{Current Progress}{\label{Progress}}

\subsection{User Submission}

While email is now, and has been for many years, available to all BaBar users, the same is not true of the Grid. Wide scale adoption in
Europe, as part of the European DataGrid and various government initiatives, has not been matched in the USA. As a result a need was 
demonstrated for a simple to install, minimal Grid client. This resulted in the BBGUtils project outlined in \ref{BBGUTILS}.

Various submission methods were considered including a web interface and a command line interface. It was rapidly ascertained that a
web interface for this project was inappropriate. The user community was uncomfortable with a move away from the command line and
discussions with the SLAC Computing Security Group rapidly outlined a number of security implications inherent in the web interface and
the need for secure data to transfer between multiple hosts.

As a result, the command line interface option was chosen. This is discussed in detail in \ref{UI}.

\subsubsection{BBGUtils}\label{BBGUTILS}

Early feedback from our user community showed that while the major computing centres such as SLAC, had adopted the Grid and have a
large number of experienced staff, many of the user institutes are not in the same position. 

From discussions with users we realised that there were two main issues which were slowing the uptake of BdbServer++. Site security
issues with the Globus Toolkit\texttrademark\cite{Globus}, especially the server components, and limited user experience in installing
Grid software software, posed a barrier to the adoption of BdbServer++.

Site security is frequently handled by different groups and individuals to those who require physics data for analysis. Their
objective, to secure their site against intrusion and limit the risk to their network, is often at odds with the users wish to perform
productive work easily. 

Many of the concerns expressed by the security groups centred around superuser based installations, potential attacks on their networks
and the ability of outside users to log into local accounts. We were able to address and avoid these problems by re-wrapping the three
client bundles of the Globus Toolkit (Data, Resource and Information) as a single package. 

This new package, which we named BBGUtils, can be installed under a standard user account and does not have the ability to allow remote
users to log into the client machine. We included copies of the various Certificate Authority Certificates and automatically install
them so the user can start submitting jobs as soon as the client is installed. 

Site specific configuration proved unnecessary as host certificates for the client machine are unnecessary and our target users
generate their personal certificates using other methods. This removes the need for this configuration and gives our users less detail
to concern themselves with.

A bash script was originally used for installation under Linux but has since evolved into a perl version which operates successfully
under Linux and Solaris.

\subsection{Request Management}

A dedicated test machine was configured at SLAC to deal with BdbServer++ requests. For security reasons this was placed outside the
SLAC firewall and is only able to access system resources specifically needed for BdbServer++.

The grid-mapfile was attached to the BaBar VO which contains all users with a valid Grid Certificate. 

The server was configured to allow it to submit jobs, under the users' ID, to the batch system for database operations.

GridFTP was enabled to allow data to be transferred to remote locations on completion. Two other file transfer applications
bbcp\cite{bbcp} and bbftp\cite{bbftp} used by BaBar that also allow third party file transfers are available, but have not been tested
yet.

\section{Implementation Issues}{\label{Results}}

\subsection{Implementation}

The majority of our implementation has been in Perl as this provides support for all platforms currently supported by BaBar. Perl is 
also widely used by members of the collaboration and has widespread use in industry and academia. 

\subsection{User Interface}\label{UI}

Development on the user interface has two tracks. Submission to SLAC is achieved by using either the globusrun or globus-job-run 
commands which come as standard with Globus. We are in the process of developing a utility to assist users in formulating their 
requests. This is intended to allow pre-prepared requests which can be used multiple times by the user.

\subsection{Server Performance}

We conducted simple tests on our development system by using a basic shell script to submit multiple simultaneous requests. The results
of these tests showed that our current server, a duel Intel\textregistered Pentium\texttrademark III-866 is able to handle in the
region of 20 simultaneous requests before we encounter dropped and lost requests. 

This is obviously not enough for a production environment but has shown us the areas we need to address to move to a production
environment. 

\subsection{User Accounts, Auditing and Caching}

We are currently looking at moving toward using a pool of user accounts for date extraction. This will have the advantage of allowing
us to catalogue what has been recently requested, and hopefully utilise cached collections to satisfy multiple requests. We will also
be able to sandbox accounts, limiting the potential damage intruders could cause.

Our progress in this area is hampered by the requirements for audit trails and the structure of the BaBar Computing Model.  As each
user is automatically assigned an account at SLAC it was a simple task for us to map the users grid certificate onto their local
account.

The use of individual user accounts is, in theory, advantageous as it allows the manipulation of data in the users home directory. In
reality this is a problem as users have only limited disk space for their accounts and with some requests this can be rapidly filled. 
With pool accounts we can allow a much greater use of disk space for a restricted number of sessions.

\section{Related Projects}{\label{Related Work}}

BdbServer++ is both middleware and a work in progress. It relies on the development of other projects to extend its capabilities and
these are evolving on a daily basis. As a result, a clear picture of how the production version of BdbServer++ will appear is difficult
to define. 

We are actively reviewing, and learning from, the work being done in the areas of distributed replica catalogues, replica location
services and automated processes for discovery and registering of data. Much work is being done on this by the EDG Work Package
2\cite{WP2} group.

Work being done by EDG Work Package 1\cite{WP1} in the area of Resource Brokers is of interest in the longer term. When coupled with a
replica catalogue this would allow us to improve performance and reliability measurably. 
  
\section{Conclusions}{\label{Conclusions}}

We have presented the current status of BdbServer++, an overview of what has lead us to this stage and an outline of the avenues we are 
currently investigating to improve it. Much work remains to be done but we have outlined a coherent strategy and discussed the issues 
we feel are important to the project and research in this area in general.

\begin{acknowledgments}

This paper is based on work supported by the UK Particle Physics and Astronomy Research Council (www.pparc.ac.uk) Studentship
PPA/S/E/2001/0338, the US Department of Energy (www.energy.gov), the Particle Physics Data Grid (www.ppdg.org) and LAPP-Annecy
CNRS/IN2P3 (www.in2p3.fr).

We wish to thank the members of the BaBar Computing Group and especially the members of the Data Distribution Group and the BaBarGrid
Group and others who have assisted in providing information and suggestions for this project. 

\end{acknowledgments}



\end{document}